\documentclass[sigconf]{acmart}
\usepackage{graphicx} 
\usepackage[ruled, lined, commentsnumbered, longend]{algorithm2e}

\usepackage{booktabs}
\usepackage{multirow}
\usepackage{longtable}
\usepackage{supertabular}
\usepackage{multicol}
\usepackage{caption}
\usepackage{subcaption}
\usepackage{graphicx}

\usepackage{newfloat}
\usepackage{listings}
\usepackage{makecell}
\usepackage{amsmath}
\usepackage{bbding}
\usepackage{colortbl}
\newcommand{\modelname}{{\textsc{ChatTutor}}}
\definecolor{mycolor2}{RGB}{249,243,243}
\definecolor{mycolor1}{RGB}{237,243,243}
\definecolor{mycolor3}{RGB}{237, 242, 245}
\definecolor{mycolor4}{RGB}{245, 245, 237}
\definecolor{mycolor5}{RGB}{245, 239, 237}

\definecolor{forestgreen}{HTML}{228B22}
\newcommand\mycheck{\textcolor{forestgreen}{\Checkmark}\xspace}
\newcommand\myx{\textcolor{red}{\XSolidBrush}\xspace}
\AtBeginDocument{%
  }

\copyrightyear{2024}
\acmYear{2024}
\setcopyright{rightsretained}
\acmConference[CIKM '24]{Proceedings of the 33rd ACM International
Conference on Information and Knowledge Management}{October 21--25,
2024}{Boise, ID, USA}
\acmBooktitle{Proceedings of the 33rd ACM International Conference on Information and Knowledge Management (CIKM '24), October 21--25, 2024, Boise, ID, USA}
\acmDOI{10.1145/3627673.3679665}
\acmISBN{979-8-4007-0436-9/24/10}

\begin{document}

\title{Empowering Private Tutoring by Chaining Large Language Models}

\author{Yulin Chen}
\authornote{equal contributions}
\email{yl-chen21@mails.tsinghua.edu.cn}
\affiliation{%
  \institution{Shenzhen International Graduate School, Tsinghua University}
  \city{Shenzhen}
  \country{China}
}

\author{Ning Ding}
\authornotemark[1]
\email{dn97@mail.tsinghua.edu.cn}
\affiliation{%
  \institution{Department of Electronic Engineering, Tsinghua University}
  \city{Beijing}
  \country{China}}

\author{Hai-Tao Zheng}
\authornote{corresponding authors}
\email{zheng.haitao@sz.tsinghua.edu.cn}
\affiliation{%
  \institution{Shenzhen International Graduate School, Tsinghua Universiy\\Pengcheng Laboratory}
  \city{Shenzhen}
  \country{China}
}

\author{Zhiyuan Liu}
\authornotemark[2]
\email{liuzy@tsinghua.edu.cn}
\affiliation{%
 \institution{DCST, BNRIST, CollegeAI, \\Tsinghua University}
  \city{Beijing}
  \country{China}}

\author{Maosong Sun}
\email{sms@tsinghua.edu.cn}
\affiliation{%
  \institution{Department of Computer Science and Technology, Tsinghua University}
  \city{Beijing}
  \country{China}}

\author{Bowen Zhou}
\email{zhoubowen@tsinghua.edu.cn}
\affiliation{%
  \institution{Department of Electronic Engineering, Tsinghua University}
  \city{Beijing}
  \country{China}}




\begin{abstract}
Artificial intelligence has been applied in various aspects of online education to facilitate teaching and learning. However, few approaches have been made towards a complete AI-powered tutoring system.
In this work, we explore the development of a full-fledged intelligent tutoring system based on large language models (LLMs). The proposed system \modelname, powered by state-of-the-art LLMs, is equipped with automatic course planning and adjusting, informative instruction, and adaptive quiz offering and evaluation.
\modelname\ is decomposed into three inter-connected core processes: \textit{interaction}, \textit{reflection}, and \textit{reaction}. Each process is implemented by chaining LLM-powered tools along with dynamically updated memory modules. 
To demonstrate the mechanism of each working module and the benefits of structured memory control and adaptive reflection, we conduct a wide range of analysis based on statistical results and user study. The analysis shows the designed processes boost system consistency and stability under long-term interaction and intentional disruptions, with up to 5\% and 20\% increase in performance respectively. 
Meanwhile, we also compare the system with scripts from real-world online learning platform and discuss the potential issues unique to LLM-based systems.
\end{abstract}




\maketitle

\begin{figure*}[!ht]
    \centering
    \includegraphics[width=0.98\textwidth]{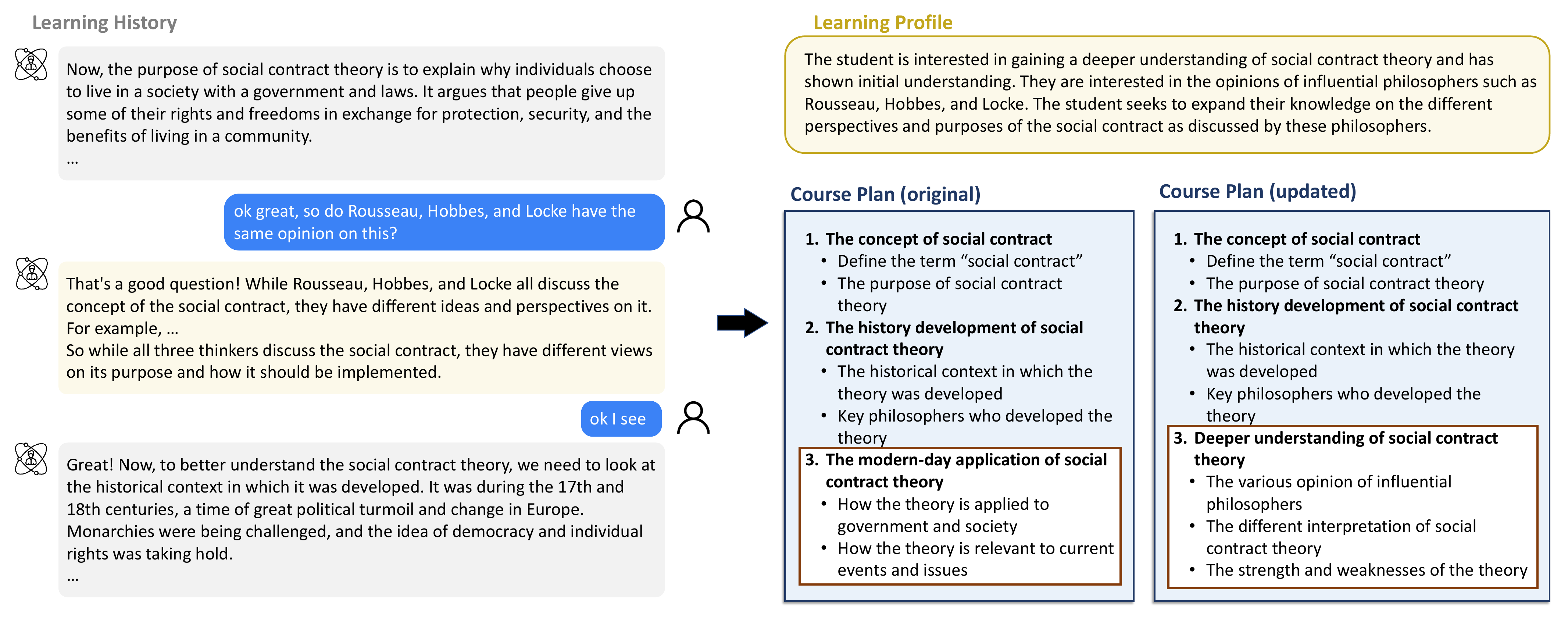}
    \caption{An example of the learning progress. The left side is the user interface directly controlled by the interaction process. The right side is the backend memory changes brought by reflection and reaction processes.}
    \label{fig:example}
\end{figure*}

\section{Introduction}

\looseness=-1 Online education, along with artificial intelligence (AI) technology, brought the aspiration of personalized tutoring within reach~\cite{bloom19842}. 
AI has been used to assist education in multiple aspects, ranging from adaptive content recommendation~\cite{costelloadaptive2009}, automatic performance evaluation ~\cite{McDonald2013AnET, Grivokostopoulou2017AnES}, to personalized instruction and dynamic feedback~\cite{Bhutoria-2022, Tang-Liang-et-al, gordonAffective2016, grawemeyer2016affect, Dzikovska2014BEETLEID}. 
Although a few early approaches have been made towards a dialogue-based intelligent tutoring system (ITS) ~\cite{Graesser_VanLehn_Rose_Jordan_Harter_2001, rusDeepTutor2014a}, most of them are domain-specific and focus primarily on guiding the users to solve a pre-defined problem.
Nevertheless, a more ultimate exploration lies in the pursuit of a full-fledged AI-driven tutoring system with greater flexibility and generalizability that teaches in a systematic and consistent manner on a much broader range of knowledge.

While previous works often employ diverse techniques jointly, including learner style classification~\cite{nihad2017analysing}, data mining~\cite{echeverria2015mirroring}, Bayesian learning~\cite{grawemeyer2016affect}, etc, the recent emergence of large language models (LLMs)~\cite{brown2020language, achiam2023gpt,touvron2023llama,team2023gemini}, like ChatGPT~\cite{openai2022chatgpt}, has broadened our imagination on new designs of intelligent tutoring systems.
LLMs impressed people firstly with the ability to generate and transform information following human instructions, then with the potential in task planning and tool usage. 
Given their extraordinary ability in providing detailed and accessible content and summarizing information, LLMs become great sources for consulting a wide range of knowledge.
Additionally, LLMs have shown remarkable capabilities in engaging in multi-turn dialogues, which makes them well-suited for tutoring applications where there is a back-and-forth exchange with a student. 
Although challenges like hallucination remain~\cite{bang2023multitask}, LLMs can adapt to expertise in specific domains and pedagogical strategies with further fine-tuning. 

In this work, we explore the potential of employing generative large language models to build a full-fledged dialogue-based personalized tutoring system.
One specialty about an ITS compared to other LLM-powered agents is that, education is a long-term cooperative process accomplished by AI and human users jointly. A well-designed tutoring system should correctly infer the human user's mental state to achieve adaptive teaching, and meanwhile, the user should be informed about the learning progress in order to cooperate more effectively. Therefore, the system faces some unique challenges in how to maintain an explainable and consistent control over the learning progress and how to adjust dynamically according to the users' response.

\modelname\ has a modularized design, encompassing three core processes-\textit{interaction}, \textit{reflection}, and \textit{reaction}, each further composed of chained LLM-powered tools to perform atomic tasks.
The processes are connected to each other through various memory modules, which store the essential data describing the overall progress and support update and retrieval. 
The design enables \textbf{structured memory control} and \textbf{adaptive reflection} on status quo. 
\modelname\ carries out every stage in education systematically and dynamically, including instructing, question answering, exercise offering and evaluating. Note that the system is designed for general purpose of learning instead of  targeting a specific subgroup.
 
Evaluation of the proposed system is conducted by analyzing statistics collected from learning logs and subjective human feedback. 
Results show that \modelname\ can satisfactorily handle various educational activities, including adaptive course plan design, consistent instructing, impromptu question answering, etc. Meanwhile ablation study demonstrates the advantage in performance stability and consistency over long-term interaction and faced with intentional disruptions. 

\section{Related Work}
Ever since the development of artificial intelligence techniques, methods and tools have been proposed to assist in teaching and learning process. 
AutoTutor~\cite{Graesser_VanLehn_Rose_Jordan_Harter_2001} is the first conversation-based intelligent tutoring system, which inspires a number of works followed~\cite{mcnamara2006improving, dmelloAutoTutor2012, Graesser2003WhyAutoTutorAT, Rus_D’Mello_Hu_Graesser_2013, vanlehn2011relative, Dzikovska2014BEETLEID}. 
In addition to AutoTutor's application to various fields, enhancement of specific aspects of education are also investigated, including adaptive feedback~\cite{Dzikovska2014BEETLEID, roscoe2013writing},
learning material recommendation~\cite{gomez2014context, mohammed2015dynamic}, 
and classifying learners~\cite{grawemeyer2016affect, nihad2017analysing, hsieh2014alg}.
Commonly adopted techniques include data mining~\cite{echeverria2015mirroring}, condition-action rule based~\cite{hsieh2014alg, gomez2014context}, and bayesian based methods~\cite{grawemeyer2016affect}, and reinforcement learning~\cite{malpani2011personalized, gordonAffective2016}.
NLP-specific techniques like semantic analysis~\cite{Graesser_VanLehn_Rose_Jordan_Harter_2001} and textual entailment~\cite{rus2006deeper, vanlehn2007developing, mccarthy2008assessing} are also adopted. 
In terms of application field, existing systems often rely on well-structured knowledge bases and therefore only target a single domain, most popular among which are health~\cite{McDonald2013AnET, el2008natural}, computer science~\cite{grivokostopoulou2013intelligent, mohammed2015dynamic}, and language learning~\cite{gomez2014context, emranSurvey2014}. 

As for applications with LLMs, 
with proper prompting and chaining, a number of works have exploited LLMs in following diverse instructions~\cite{ouyang2022training,chiang2023vicuna,ding2023enhancing,cui2023ultrafeedback}, decomposing tasks~\cite{wei2023chainofthought}, refining answers~\cite{madaan2023selfrefine, shinn2023reflexion}, using external tools~\cite{Qin2023ToolLW}, and simulating human behaviors~\cite{park2023generative}. 
While our work focuses on building an interactive tutoring system that works cooperatively with human users, featuring dynamic reflection.

\begin{figure*}[!ht]
    \centering
    \includegraphics[width=0.96\textwidth]{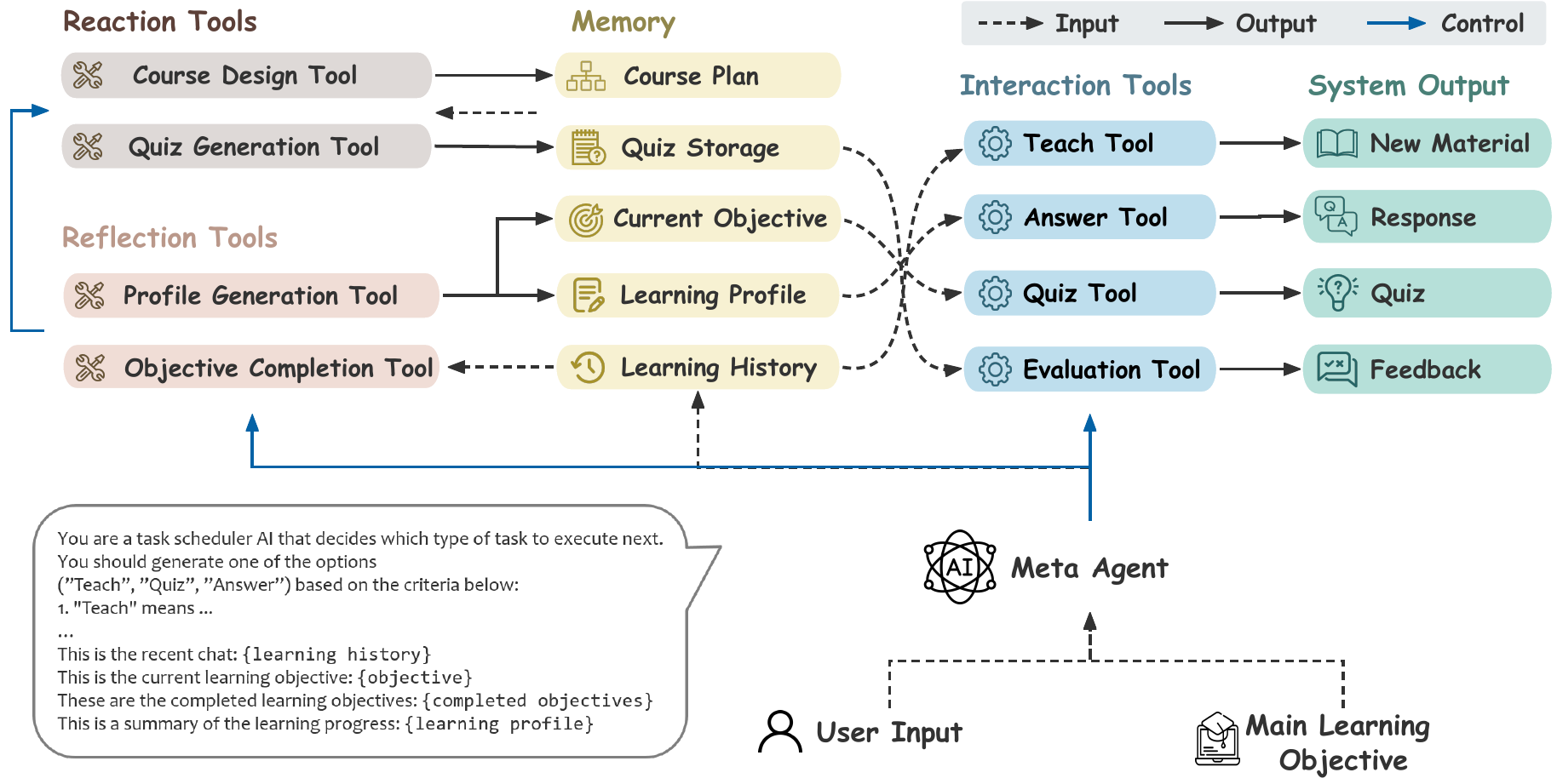}
    \caption{An overview of the system's modular implementation and execution in a single round of conversation.}
    \label{fig:overview}
\end{figure*}

\section{Overview of \modelname}

\modelname\ is essentially a dialogue-based tutoring system that aims to help learners acquire knowledge on one given topic systematically. As shown in Figure~\ref{fig:example}, the whole learning process is carried out in natural language conversations, with time-to-time backend reflection and reaction to update the memories. 
This section gives a general picture of the system's workflow. We start with explaining the design principles by introducing three underlying processes within the system. Then, we briefly go over the components employed to realize each process. Finally, we provide a complete introduction to how each process and components work together.
Note that the proposed system mainly explores autonomous tutoring and adaptive system reflection with chained LLMs, while we do acknowledge the potential of fallacious and biased output due to inherent flaws in LLMs.

\subsection{Design Principles}
The system design demonstrates the breakdown into three core system processes: \textit{Interaction}, \textit{Reflection}, and \textit{Reaction}. They each has a modularized implementation and is connected to one another to form an execution loop that empowers the whole education process. 

\noindent \textbf{Interaction.} The interactive dialogue between the system and the user is the media for tutoring and learning, and therefore is the major process of the designed system. LLMs like ChatGPT can interact with users in a responsive and robust way in daily chit-chat. However, tasking them with long-term purposeful interaction is still tricky given the restriction on context length.
As for educational purposes, it is especially important to keep the interaction on track, meanwhile ensuring its accessibility and informativeness. 

\noindent \textbf{Reflection.} To facilitate interaction, we devise a reflection process to generate high-level insights on the learning progress, which serves as global information~\cite{park2023generative} input into the system module. It is expected to help adjust system response dynamically based on user preference and behavior to achieve personalized tutoring.

\noindent \textbf{Reaction.} Along with reflection, reaction refers to the automatically triggered system behavior afterward, including adjustment of course plan and quiz generation. It differs from the interaction process in that \textit{interaction} is always triggered directly by a new round of response, while \textit{reaction} is performed at the backend from time to time, subsequent to the reflection process.

\subsection{Components} 
The introduced three processes are further realized by separate components that support or execute a single task. There are three kinds of components: \textit{Tools}, \textit{Memories}, and \textit{Meta Agent}. 

\noindent \textbf{Tools.} Under the principled design, each process is embodied by a set of tasks performed either sequentially or in parallel. 
For instance, there are diverse ways of engaging with the student, such as providing instructions, addressing questions, administering quizzes, and offering feedback. This variation in approaches complicates the development of a single unified solution.
We therefore devise separate modules for each specific task to ensure performance. We term those modules as ``tools'', and that each tool is a task-specific prompted LLM responsible for generating system output or updating memories, as shown in Figure~\ref{fig:overview}. 
For example, \textit{interaction} is broken down into four types of response in terms of education function, each hosted by one well-prompted tool. At each round, only one tool is used to generate the response. 

\noindent \textbf{Memories.}
Apart from tools, data storage is required to host information generated by reflection and reaction processes, while also supporting querying and updating.
We propose four types of memories to record the progress and current status of learning, each stored in distinct data format and supports different ways of querying and updating. Another critical feature of the memories is that they serve as a linkage between different sets of tools to pass on information to control tool output.
The detailed description of each tool and memory can be found in the next section.

\noindent \textbf{Meta Agent.} 
Above all three processes, we introduce meta agent, the single access of the control flow. It is powered by LLM and prompted to decide what specific tasks to execute next. See Figure~\ref{fig:overview} for an example prompt for controlling the interaction process. The template contains helpful information retrieved from the memory and asks for an output deciding the type of interaction process. 
In our implementation, the meta agent only controls the interaction tools, while we set a fixed time interval for the execution of reflection process.


\begin{figure*}[!htbp]
    \centering
    \includegraphics[width=0.98\textwidth]{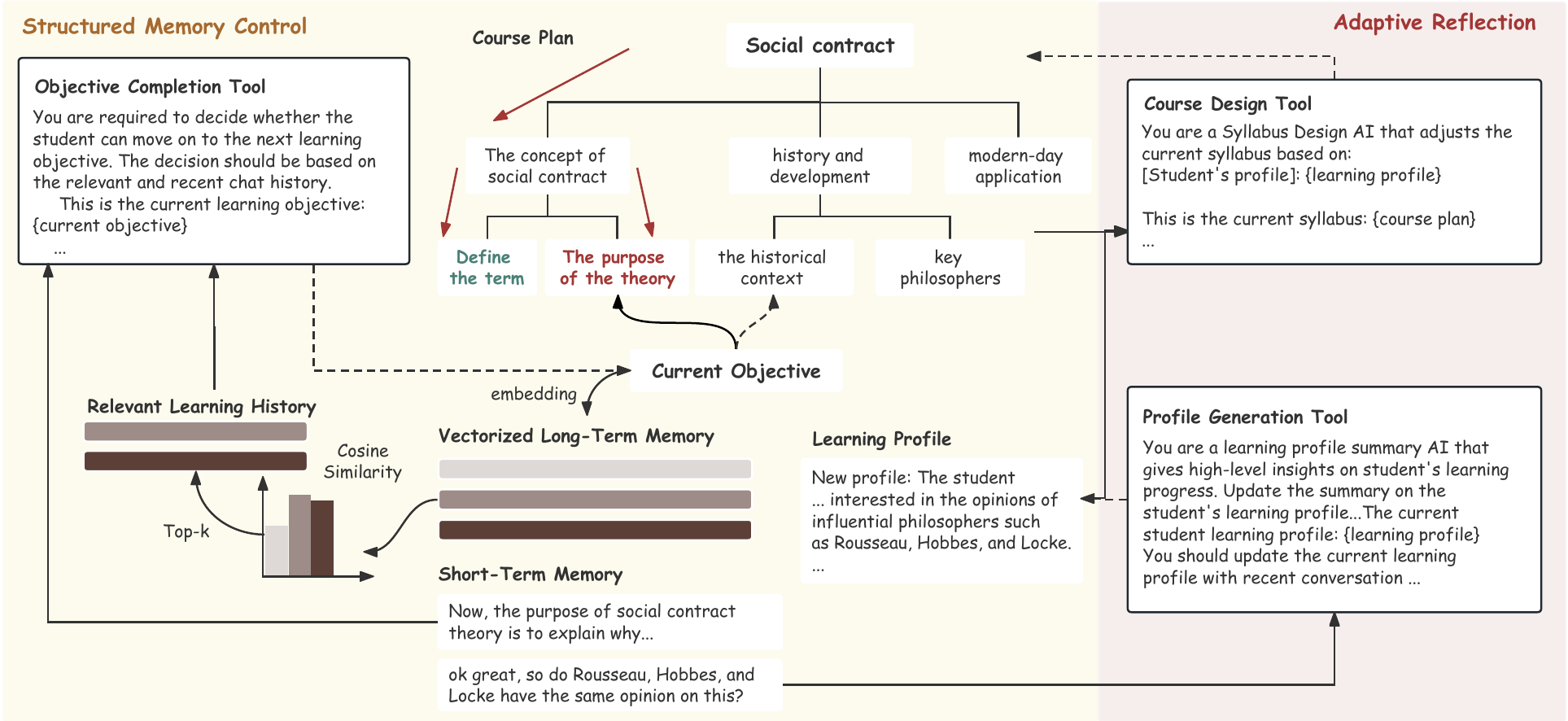}
    \caption{A detailed illustration of how course plan is stored and manipulated structurally and how reflection process helps customize the reaction followed.}
    \label{fig:memory}
\end{figure*}

\begin{table*}[!htbp]
    \centering
    \caption{A summary of the detailed tool usage. ``Input'' means the memories are part of the prompt.}
    \scalebox{0.84}{
    \begin{tabular}{lllll}
    \toprule
        \textbf{Process} & \textbf{Tool Name} & \textbf{Execution Condition} & \textbf{Input} & \textbf{Output/Update} \\
    \midrule
        \cellcolor{mycolor3} & Teach Tool & Meta agent & Learning history, Current objective, Learning profile & System output \\
        \cellcolor{mycolor3} & Answer Tool & Meta agent & Learning history & System output \\
        \cellcolor{mycolor3} & Quiz Tool & Meta agent & Quiz storage, Learning profile & System output  \\
        \multirow{-4}{*}{\cellcolor{mycolor3} \textbf{Interaction}} & Evaluation Tool & Quiz & Learning history & System output \\
        \midrule
        \cellcolor{mycolor2} & Profile Generation & Each round & Learning history, Learning profile & Memory: Learning profile \\
        \multirow{-2}{*}{\cellcolor{mycolor2} \textbf{Reflection}} & Objective Completion & Each round & Learning history, Current objective & Memory: Current objective \\
        \midrule
        \cellcolor{mycolor5} & Course Design &  Profile generation & Course plan, Learning profile & Memory: Course plan \\
        \multirow{-2}{*}{\cellcolor{mycolor5} \textbf{Reaction}} & Quiz Generation & Objective completion & Learning history, Current objective & Memory: Quiz storage \\        
    \bottomrule
    \end{tabular}}
    \label{tab:tools}
\end{table*}

\subsection{Overview of Control Flow}

Above all, all designs serve for the ultimate goal of better interaction with the users. 
The system reflects from time to time to update cognition on the overall progress, and in turn refines the interaction production with new insights.
At the frontend, the user first inputs what to learn with desired difficulty level. Then the system automatically calls the course design tool to generate the initial course plan, and starts the conversation accordingly.
Upon receiving a new round of user input, the meta agent decides which interaction tool to use and the tool executes the task correspondingly to generate a new response with queried information from memories. 
At the backend, the reflection tools are triggered to reflect on the status quo and update the learning profile and current objective, after which the reaction tools will be triggered immediately to generate new quiz questions and update the course plan.

As shown in Figure~\ref{fig:overview}, the right side represents the interaction process that is presented on the user interface, while the left side demonstrates the backend processes that are responsible for generating and updating memory modules.
Practically, throughout each dialogue session, the reflection and reaction processes run alternatively at the backend, where the output result is periodically utilized by the interaction process to produce the final response to the user in each round. 
Table~\ref{tab:tools} presents detailed usage of each tool in the three processes, including the input and output memory content, and the condition for tool execution.
The learning proceeds until all objectives in the course plan have been completed.


\section{Structured Control and Adaptive Reflection}
As described in the previous section, the system functions through the combination of memory modules and LLM-powered tools, where memories are extracted as part of prompt in the tools. Table~\ref{tab:tools} presents detailed usage of each tool in the three processes, including the input and output memory content, and the condition for tool execution.
We further describe the key features of \modelname\ along with explaining the functionality of core components below.

\subsection{Structured Memory Control}

The interactive and cooperative feature of a tutoring system calls for the need to communicate with the users effectively about current and future progress. Meanwhile, it is also important to keep the system itself aware of the progress to ensure better stability. We therefore design various memory modules in different storage format and function to support the mutual communication.

\noindent \textbf{Course Plan.} The course plan is stored in a tree structure, with each node representing an atomic topic in the course, and its child nodes representing the sub-topics. The course is expected to be taught and learnt in depth-first traverse order. Current objective is a pointer pointing to the next uncompleted objective node in the tree to denote current progress. 
Such structure allows for presentation to the users, informing them of the overall status of learning, while enabling mechanistic operation by the system.
Specifically, the course design tool is used at the beginning of the learning to generate the initial course plan based on user's desired topic and difficulty level. In each new round of conversation, objective completion tool is called to update status of the current objective based on the recent and relevant learning history. Then, the course design tool is asked to update the current course plan while maintaining the completed objectives. 

\noindent \textbf{Learning History.} As for learning histories, the recent history is stored as plain text that can be directly fed into the LLM, whereas the relevant history is stored along with their embedding and queried with cosine similarity with embeddings of current objective upon usage. The detailed mechanism can be seen in Figure~\ref{fig:memory}.
Meanwhile, the benefit of explicitly collecting completed objectives also extends to more effective quiz offerings, which will be detailed in the next section.

\noindent \textbf{Quiz Storage.} A crucial function of an ITS system is to offer adaptive quiz that helps the learner review and master what has been learned. In \modelname, the LLM is instructed to generate quiz questions based on learning materials and formatted as a structured json string. The questions are stored corresponding to each learning objective with explicit status marking, and extracted in order whenever the quiz tool is called. The quiz questions will keep appearing in the next quiz batch until it is answered correctly by the learner.



\subsection{Adaptive Reflection and Reaction}
\looseness=-1 Reflection and reaction processes at the backend are closely bound to each other in \modelname, whereas reflection process generates high-level insight about the learning progress, reaction process updates the structured memory modules based on the insights. The sequential and dependent design could more accurately infer about the status and enhance the stability and adaptiveness of the system behavior.

\noindent \textbf{Learning Progress Control.}
One core function of reflection process is to control the learning progress by determining when to move on to next learning objective. Objective completion tool is prompted to judge whether the current objective has been completed based on queried learning histories based on text embedding similarity. 
Whenever the current objective is considered completed, the status of the course plan and the pointer will automatically be updated. Meanwhile, the reaction process ``quiz generation'' will be triggered as well. It is prompted to generate several representative quiz questions for the completed current objective, with relevant queried learning history provided, which ensures the stability and relevance of the generated quiz question. The questions are stored in the memory until the meta agent decides it is time for a quiz, where the corresponding quiz questions are retrieved from the storage for the completed objectives and further filtered and organized by the quiz tool in the interaction process to present to the user.

\noindent \textbf{Profile and Course Plan Update.}
Apart from reflecting on objective status, an important component is user's learning profile. Learning profile summarizes what the user has learned and gives high-level insight on the user's preference based on conversation history. Though not directly presented to users, it is crucial to the stability of system's memory update and overall understanding of the learning process. It is especially useful as part of the input to course design tool to provide direction for course plan adjustment.

At each round of conversation, the system automatically reflects on the recent dialogue and updates learning profile with profile generation tool. The tool is a prompted LLM that takes recent dialogues and current profile summary as input and outputs a new version of learning profile, summarizing the learned knowledge, the user's reaction and preference mainly. Then it is fed into the course design tool for a new version of course plan generation. Figure \ref{fig:memory} provides an example of profile generation tool generating high-level insight of \textit{``the student seeks to expand their knowledge on the different perspectives and purposes of the social contract as discussed by the philosophers.''} after the user asks a follow-up question about different philosophers' opinion. This further leads to an updated course plan that enhances deeper understanding of the theory.

\section{Experiments}

To demonstrate and analyze the features of our tutoring system, the experiments are conducted in two folds. We invite a number of users to learn a series of pre-defined topics using the system. During interaction, we collect critical statistics and record the conversation for future analysis. After learning completes, the users are required to answer a questionnaire to rate their experience with the system from multiple perspectives. We also develop ablation systems to better understand the effect of each process and module.

\begin{figure*}[!htb]
    \centering
    \includegraphics[width=0.9\textwidth]{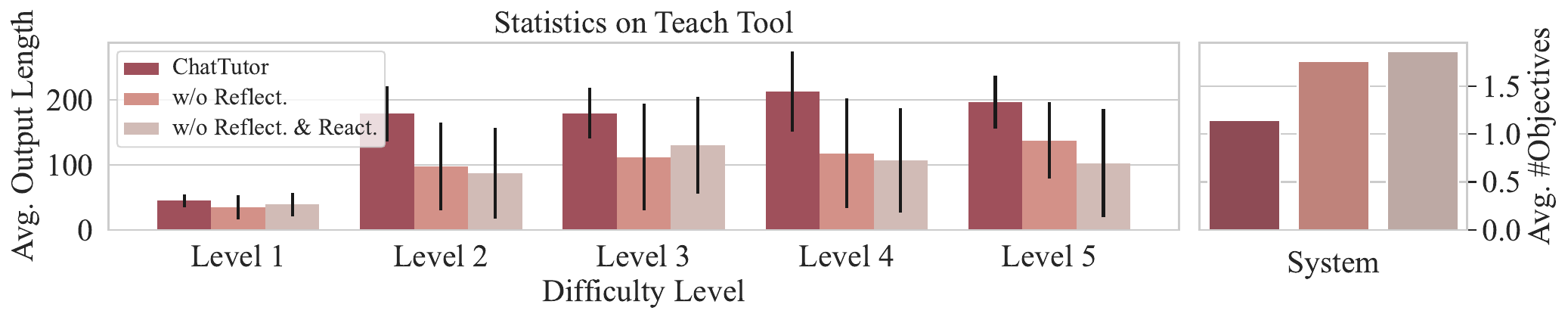}
    \caption{Average output length (calculated by the number of words) and the number of objectives covered in each output for different systems. Average number of objectives are manually annotated with 50 randomly sampled response from each system.}
    \label{fig:teach_tool}
\end{figure*}

\begin{figure*}[!htb]
    \centering
    \includegraphics[width=0.9\textwidth]{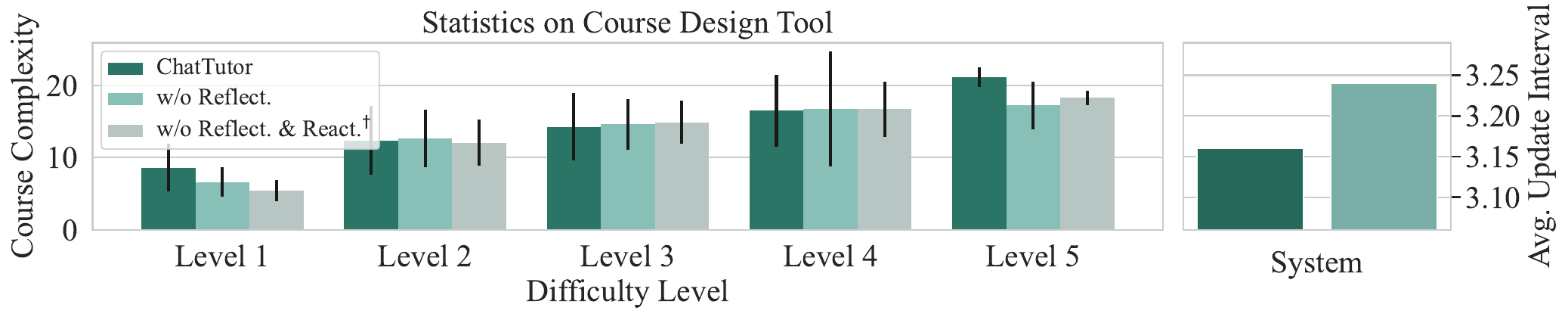}
    \caption{Average course plan complexity (calculated by the number of objectives) and update interval (calculated by the number of conversation rounds in between) by course design tool for different systems.$\dag$ means this is the baseline statistics as the system without reflection or reaction processes has a fixed course plan throughout learning.}
    \label{fig:course_plan}
\end{figure*}

\subsection{Experimental Design}

\noindent \textbf{System Setup.} In addition to the main system, we implement two ablation systems with only partial functions. Specifically, we have one system without reflection process (w/o Reflection) and another with both reflection and reaction processes removed (w/o  Reflection \& Reaction). 
For \modelname\ w/o Reflection, the reflection process is removed so no learning profile is generated throughout the whole process, and the reaction process is triggered at a fixed time interval with limited input information.
For \modelname\ w/o Reflection \& Reaction, only recent learning history and the initial course plan are available for tools.

\noindent \textbf{Main Learning Objectives.} For system evaluation, we collect learning objectives that cover a wide range of academic subjects and some daily life phenomena, varying in granularity and language format.
We first ask GPT-4 to generate a set of general academic domain. Then we ask for generation of more fine-grained subjects under each domain and the related classic concepts. 
Besides, we also include some daily phenomena that may inspire people's wondering. We encourage GPT-4 to generate a typical list of them in diverse language style. 
To demonstrate the system's robustness in dealing with various types of learning objectives, we randomly sample 80 topics from generated fine-grained subjects, atomic concepts, and daily wondering. To make the learning process more diverse and controllable, we also design 5 difficulty levels according to Bloom's taxonomy~\cite{bloom2020taxonomy} and randomly assign them to each topic. 
In evaluation, each topic is learned independently with three systems, making up altogether 240 courses.
Table~\ref{tab:sample_topics} shows a sample of learning objectives we adopt.

\begin{table}[!htp]
    \centering
    \caption{Examples of learning objectives used for evaluation.}
    \scalebox{0.72}{
    \begin{tabular}{c|l}
        \toprule
       \textbf{Category} & \textbf{Main Learning Objective} \\
       \midrule
        \multirow{3}{*}{Subjects} & Developmental psychology \\
        & Impressionism \\
        & Computer architecture \\
        \midrule
        \multirow{2}{*}{Atomic Concepts} & Stream of consciousness \\
        & Earth's mantle \\
        \midrule
        \multirow{2}{*}{Daily Wondering} & How do bees communicate and find their way back to the hive? \\
        & How do rainbows form and why do they have different colors? \\
        \bottomrule
    \end{tabular}}
    \label{tab:sample_topics}
\end{table}

\noindent \textbf{Participants.}
We invite 13 average adult users who are proficient in English to participate in learning. Every single course is randomly assigned to one user, while we make sure that each participant does not get repeated course topics.


\looseness=-1 \noindent \textbf{Statistical Analysis.}
We collect various statistics for analysis, including (1) \textit{Complexity of course plan} reflects the ability to design adaptive course plan; (2) \textit{Average length of system response} and \textit{Average number of objectives per response} are indicators of instruction informativeness; (3) \textit{Frequency of course plan update} shows the reflective feature of the system; and (4)  \textit{Frequency of in-course quiz} explores the pattern of quiz offerings.

\noindent \textbf{Survey Design.}
After completing the course, the learner is required to answer a survey composed of 8 questions targeting different aspects of the system. Each question is a statement to be rated on a 1$\sim$5 scale, where higher scores mean better agreement with the statement.
Table~\ref{tab:survey_q} presents the statements by category.

\begin{table*}[!tb]
    \centering
    \caption{Survey results for learning courses at difficulty level 1$\sim$3. $\dag$ means the score evaluates the initial course plan only, as no changes in course plan happen throughout the learning process. It could be viewed as the static quality evaluation of course plan generated from scratch. $^*$ means p-value $<$ 0.1 using t-test.}
    \scalebox{0.9}{
    \begin{tabular}{l|cc|cc|cc|cc}
    \toprule
       \multirow{2}{*}{\textbf{System}}  &  \multicolumn{2}{c|}{\cellcolor{mycolor1} \textbf{Course Plan}} & \multicolumn{2}{c|}{\cellcolor{mycolor2}\textbf{Instruction}} & \multicolumn{2}{c|}{\cellcolor{mycolor3}\textbf{Question Answering}} & \multicolumn{2}{c}{\cellcolor{mycolor4}\textbf{Quiz}} \\
         & \cellcolor{mycolor1}Relevance & \cellcolor{mycolor1}Coherence & \cellcolor{mycolor2}Consistency & \cellcolor{mycolor2}Accessibility & \cellcolor{mycolor3}Timeliness & \cellcolor{mycolor3}Consistency & \cellcolor{mycolor4}Relevance & \cellcolor{mycolor4}Judgment \\
         \midrule
         
         \modelname & \textbf{4.72} & 4.51 & 4.32 & \textbf{4.77}$^{*}$ & 4.41 & 4.82  & \textbf{4.88} & 4.24 \\
         \modelname\ w/o Reflect. & 4.71 & \textbf{4.62} & \textbf{4.46} & 4.66 & 4.64 & 4.85  & 4.75 & \textbf{4.65} \\
         \modelname\ w/o Reflect. \& React. & 4.97$^\dag$ & 4.77$^\dag$ & 4.34 & \textbf{4.77} & \textbf{4.75} & \textbf{4.95}  & 4.86 & 4.36 \\
         
         \bottomrule
    \end{tabular}}
    
    \label{tab:survey_easy}
\end{table*}

\begin{table*}[!tb]
    \centering
    \caption{Survey results for learning courses at difficulty level 4$\sim$5. $\dag$ means the score evaluates the initial course plan only, as no changes in course plan happen throughout the learning process. It could be viewed as the static quality evaluation of course plan generated from scratch. $^*$ and $^{**}$ means p-value $<$ 0.1 and $<$ 0.05 using t-test.}
    \scalebox{0.9}{
    \begin{tabular}{l|cc|cc|cc|cc}
    \toprule
       \multirow{2}{*}{\textbf{System}}  &  \multicolumn{2}{c|}{\cellcolor{mycolor1} \textbf{Course Plan}} & \multicolumn{2}{c|}{\cellcolor{mycolor2}\textbf{Instruction}} & \multicolumn{2}{c|}{\cellcolor{mycolor3}\textbf{Question Answering}} & \multicolumn{2}{c}{\cellcolor{mycolor4}\textbf{Quiz}} \\
         & \cellcolor{mycolor1}Relevance & \cellcolor{mycolor1}Coherence  & \cellcolor{mycolor2}Consistency & \cellcolor{mycolor2}Accessibility & \cellcolor{mycolor3}Timeliness & \cellcolor{mycolor3}Consistency & \cellcolor{mycolor4}Relevance & \cellcolor{mycolor4}Judgment \\
         \midrule
         \modelname & \textbf{4.87}$^{*}$ & \textbf{4.87}$^{**}$ & \textbf{4.26}$^{**}$ & 4.53 & 3.67 & \textbf{5.00} & 4.87 & \textbf{4.20} \\
         \modelname\ w/o Reflect. & 4.67 & 4.60  & 4.14 & 4.87 & 4.80 & 4.67 & 4.40 & 4.13 \\
         \modelname\ w/o Reflect. \& React & 4.93$^\dag$ & 4.73$^\dag$  & 3.73 & \textbf{5.00} & \textbf{4.33} & 4.93 & \textbf{4.93} & 4.00 \\
         \bottomrule
    \end{tabular}}
    
    \label{tab:survey_difficult}
\end{table*}

\noindent \textbf{Stability Analysis.} To further demonstrate the benefit of our system design, intentional disruption to the learning process is conducted to test the stability of the system. We take 15 most difficult topics (difficulty 4$\sim$5) in our list and for each learning process apply 3 consecutive rounds of attacks with ChatGPT generating a random question. The system is expected to answer robustly to the question and resume the original learning course after the disruption. we manually annotate the quality of resumed learning after disruption and the quality of response for the attack questions. 
In our experiments, each topic is learned independently with three systems. We evaluate each learning process in terms of 1) \textit{Repeat}: the degree of repetition in course material, 2) \textit{Omit}: whether there is omission of sub-topics while instructing, and  3) \textit{Response}: whether the system responds robustly to user's random questions. 
Each learning is scored with 3 aspects respectively and the score can be 0, 0.5 or 1. Specifically, for response robustness, direct ignorance of the question or repeated template answer like ``Let's stay focused on the course material.'' will be considered a sign of lack of robustness.

\begin{table}[!t]
    \centering
    \caption{The complete survey questions. Learners are asked to rate the compatibility of each statement on a scale of 1$\sim$5.}
    \scalebox{0.8}{
    \begin{tabular}{l}
    \toprule
        \textit{Course Plan} \\
        1. Relevance: The course plan is relevant to the main objective. \\
        2. Coherence: The course plan is coherent and logical. \\
         \midrule
         \textit{Instruction} \\
         3. Consistency: The instruction content strictly follows the course plan. \\
         4. Accessibility: The language used is easy to understand. \\
         \midrule
         \textit{Question Answering} \\
         5. Timeliness: The learner's questions always get immediate response. \\
         6. Consistency: The response is consistent with learning material. \\
         \midrule
         \textit{Quiz} \\
         7. Relevance: The quiz questions match what has been covered. \\
         8. Judgment: The quiz evaluation is accurate in parsing and scoring. \\
         \bottomrule
    \end{tabular}}
    \label{tab:survey_q}
\end{table}

\subsection{Results}
\noindent \textbf{Statistical Results.}
Figure~\ref{fig:teach_tool} presents the statistical characteristics related to teach tool, including average length of output and the average number of objectives covered in each generation. Overall all systems can generate tailored output according to difficulty level. Higher difficulty comes along with longer and more informative output. It means the teach tool is successfully aware of the dynamic prompting controlled by difficulty.
What is worth noting is that \modelname\ generates significantly longer output with the smallest variation. It demonstrates that \modelname\ is able to consistently generate informative content on the given topic, which is further testified by the number of objectives covered in each output. 
This phenomenon shows the benefits of structured memory control, where the objective completion tool reflects on and updates the current objective so that the teach tool is prompted to give new materials on one specific atomic objective.

Figure~\ref{fig:course_plan} showcases the behavior of course design tool. Note that all systems start with the same initial course plan. The system without reflection relies on learning history in course plan updates, while \modelname\ takes advantage of the learning profile. All systems can generate difficulty-aware course plans, seen in a positive correlation between complexity and difficulty. 
Note that \modelname\ shows more stability in complexity control compared to ablation systems without intermediate learning profile generation step, demonstrating the effectiveness of reflection process stabilizing the reaction process.
In terms of frequency of updates, we observe that \modelname\ seems more prone to making updates, meaning that it is easier to detect helpful information in a compressed learning profile compared to verbose learning history. 

As for in-course quizzes, the average interval between quizzes for \modelname\ and ablation system are 6.91 and 10.20, respectively. Without the backend processes, the meta agent has little information available to suggest effective quiz generation. 

\begin{figure*}[!t]
    \centering
    \includegraphics[width=0.95\textwidth]{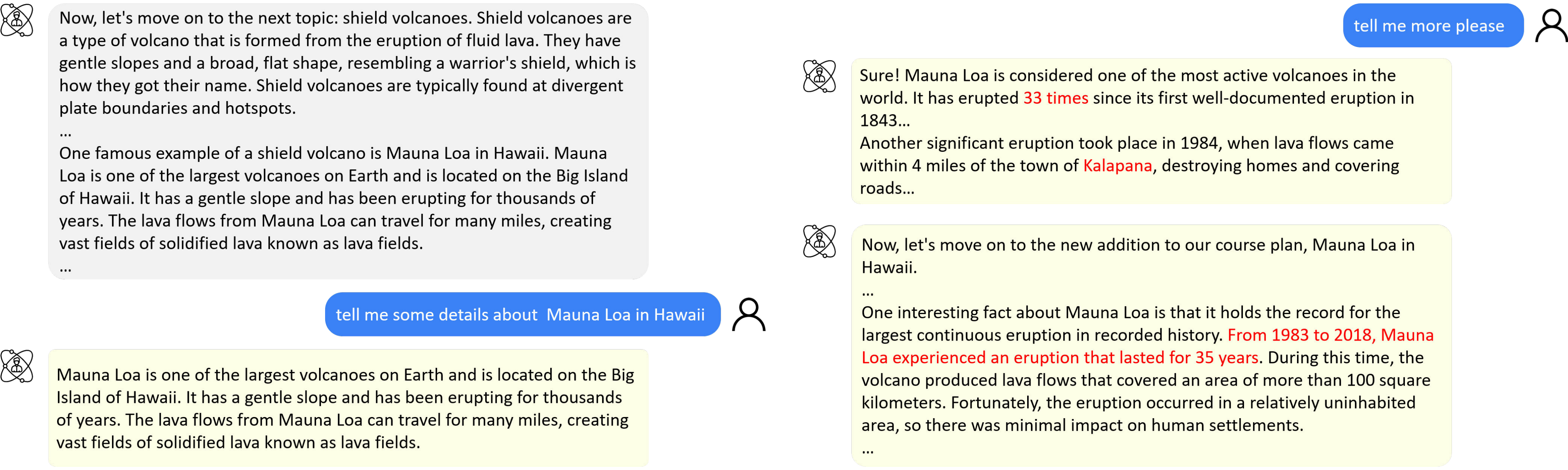}
    \caption{An example of hallucination. Hallucinated generation as been colored \color{red}{red}.}
    \label{fig:case_vol}
\end{figure*}

\noindent \textbf{Survey Results.}
Table~\ref{tab:survey_easy} and Table~\ref{tab:survey_difficult} present results on survey questions. Overall, with powerful ChatGPT, all systems demonstrate promising usability and quality.
The designed reflection and reaction processes offer advantages in complex learning settings. For intricate course plan updates, the main system achieves better coherence. Through reflection and reaction, the system maintains better control with much more consistent instructions and more faithful quiz evaluation during longer conversations. 
While all three systems seem to generate highly relevant quiz questions, \modelname\ actually generates much more fine-grained questions querying about detail information in the learning material, while the questions from ablation systems are largely answerable based on the course plan, as demonstrated in Table~\ref{tab:case_course_plan}. This advantage is attributed to online quiz generation and storage unique to reflection and reaction processes.
However, it should also be noted that the timeliness in response of \modelname\ is compromised by backend processes. It indicates that prompts with global information may interfere with the LLM's ability to focus locally and generate timely and coherent response.

\noindent \textbf{Stability Analysis.} As shown in Table~\ref{tab:stability}, thanks to the reflection and reaction processes, \modelname\ has the overall best performance in terms of stability and robustness, with up to 20\% increase compared to ablation systems. Whereas there is a clear trade-off between the repetition and omission of course material in the two ablation systems, while both signify instability. The fact that \modelname\ tends to ignore user's irrelevant questions more often also echoes finding in user study and highlights the reconciling effect between robustness and stability and controllability.

\begin{table}[!tbp]
    \centering
    \caption{Results of stability evaluation. ``Repeat'' means the repetition of learning materials and ``omit'' means some topics are skipped. Higher score indicates better stability.}
    \scalebox{0.95}{
    \begin{tabular}{l|cc|c|c}
    \toprule
        \textbf{System} & \textbf{Repeat} & \textbf{Omit} & \textbf{Response} & \textbf{Overall} \\ \midrule
        \modelname & 0.50 & 0.93 & 0.60 & \textbf{0.68} \\
        w/o Reflect. & 0.60 & 0.47 & 0.40 &  0.49 \\
        w/o Reflect. \& React. & 0.33 & 0.67 & 0.67 &  0.56 \\ 
         \bottomrule
    \end{tabular}}
    \label{tab:stability}
\end{table}

\subsection{Case Study}

In this section, we further demonstrate how \modelname\ behaves with detailed case studies. 
To compare with real-world education scenario, we adopt the machine learning course on Coursera platform. As show in Table~\ref{tab:case_course_plan}, it can be seen that \modelname\ can satisfactorily cover the major topics of the course, meanwhile also maintaining a logical dependence between crucial concepts. On the other hand, it should be noted that while \modelname\ tends to propose a wide range of concepts, real world teaching pays more attention to technical problems, including how to solve a specific machine learning problem and what practical tricks are commonly used. This down-side could be compensated by the adaptiveness and timely responsiveness of \modelname, where users can motivate more in-depth discussion with impromptu questions.

While \modelname\ largely provides accurate information on ``machine learning'', in another case featuring ``volcanoes'', we find that the system stumbles when users ask for more details. For example, as show in Figure~\ref{fig:case_vol}, the system makes typical hallucination due to the knowledge cutoff of training data, and also confuses two towns in Hawaii in a specific eruption. When being pushed to provide more information, it hallucinates the eruption duration as well. Problems like this could be mitigated with retrieval-augmented generation technique given a relevant knowledge base. 

\begin{table*}[!ht]
    \centering
    \caption{Example quiz questions generated by different systems on the topic of ``Gravity Waves''. ``Fine-grained'' means the questions are more  detailed so that the answer is not obvious from the course plan.}
    \scalebox{0.8}{
    \begin{tabular}{p{9cm}<{\centering}|p{9cm}p{2cm}<{\centering}}
    \toprule
        \textbf{Course Plan} & \textbf{Quiz Questions} & \textbf{Fine-grained?} \\
        \midrule
        \multirow{4}{*}{
            \makecell[l]{1. Introduction to Gravity Waves\\
	\ \ a. Definition and Key Concepts\\
		\ \ \ \ i. Differentiation between Gravity Waves and Gravitational Waves\\
		\ \ \ \ ii. Causes and Influences of Gravity Waves\\
	\ \ b. Factors Affecting Gravity Wave Formation\\
		\ \ \ \ i. Atmospheric Stability\\
		\ \ \ \ ii. Wind Shear\\
		\ \ \ \ iii. Topography and Surface Obstacles\\
  \\
2. Characteristics and Properties of Gravity Waves\\
	\ \ a. Wave Amplitude and Wavelength\\
		\ \ \ \ i. criteria for wave breaking\\
		\ \ \ \ ii. energy transfer and propagation\\
	\ \ b. Wave Speed and Frequency\\
		\ \ \ \ i. dispersion relation\\
		\ \ \ \ ii. role of buoyancy and ambient fluid properties\\
  \\
3. Observation and Detection of Gravity Waves\\
	\ \ a. Remote Sensing Techniques\\
		\ \ \ \ i. satellite imagery\\
		\ \ \ \ ii. lidar and radar\\
 \ \ ...
  }
        } &
        \multicolumn{2}{c}{\textbf{\modelname}}  \\
        \cline{2-3}
        & \makecell[l]{1. What is amplitude in the context of gravity waves? \\
a) The distance between successive crests or troughs of a wave \\
b) The maximum displacement of particles within a wave \\
c) The rate at which wave energy is transferred vertically \\
d) The disrupted, turbulent state of a wave} & \mycheck \\
\cline{2-3}
& \makecell[l]{2. How does wavelength influence the behavior of gravity waves? \\
a) It determines the spatial scale of the wave \\
b) It represents the distance over which the wave repeats itself \\
c) It influences the rate at which wave energy is transferred vertically \\
d) It determines the amplitude of the wave} & \mycheck \\
\cline{2-3}
\cline{2-3}
& \multicolumn{2}{c}{\textbf{\modelname\ w/o Reflect. \& React.}}  \\
\cline{2-3}
& \makecell[l]{1. What is the main focus of the course "Introduction to Gravity Waves"?\\
a) Gravitational Waves\\
b) Atmospheric Stability\\
c) Causes and Influences of Gravity Waves\\
d) Topography and Surface Obstacles} &  \myx \\
\cline{2-3}
& \makecell[l]{2. What are the factors affecting gravity wave formation?\\
a) Wave Speed and Frequency\\
b) Wind Shear\\
c) Wave Amplitude and Wavelength\\
d) Dispersion Relation} & \myx \\
\bottomrule
\end{tabular}}
    \label{tab:quiz}
\end{table*}

\begin{table*}[]
    \centering
    \caption{Course plan comparison for ``Machine Learning''.}
    \scalebox{0.85}{
    \begin{tabular}{l|l}
    \toprule
        \textbf{Coursera} & \textbf{\modelname}\\
        \midrule
        \makecell[l]{
        1. introduction\\ 
\ \ \ \ a. welcome to machine learning\\ 
\ \ \ \ b. supervised learning\\ 
\ \ \ \ c. unsupervised learning\\ 
2. linear regression with one variable\\ 
\ \ \ \ a. model representation\\ 
\ \ \ \ b. cost function\\ 
\ \ \ \ c. gradient descent\\ 
\ \ \ \ d. gradient descent for linear regression\\ 
3. linear algebra review\\ 
\ \ \ \ a. matrices and vectors\\ 
\ \ \ \ b. addition and scalar multiplication\\ 
\ \ \ \ c. matrix vector multiplication\\ 
\ \ \ \ d. matrix matrix multiplication\\ 
\ \ \ \ e. matrix multiplication properties\\ 
\ \ \ \ f. inverse and transpose\\ 
4. linear regression with multiple variables\\ 
\ \ \ \ a. multiple features\\ 
\ \ \ \ b. gradient descent for multiple variables\\ 
\ \ \ \ c. gradient descent in practice i feature scaling\\ 
\ \ \ \ d. gradient descent in practice ii learning rate\\ 
\ \ \ \ e. features and polynomial regression\\ 
\ \ \ \ f. normal equation\\ 
\ \ \ \ g. normal equation noninvertibility\\ 
\ \ \ \ h. working on and submitting programming assignments\\ 
5. octave matlab tutorial\\ 
\ \ \ \ a. basic operations\\ 
\ \ \ \ b. moving data around\\ 
\ \ \ \ c. computing on data\\ 
\ \ \ \ d. plotting data\\ 
\ \ \ \ e. control statements for while if statement\\ 
...
        } & \makecell[l]{
         1. Introduction to Machine Learning\\ 
\ \ \ \ \ \ \ \ a. Definition of Machine Learning\\ 
\ \ \ \ \ \ \ \ b. Importance and Applications of Machine Learning\\ 
\ \ \ \ \ \ \ \ c. Types of Machine Learning\\ 
2. Supervised Learning\\ 
\ \ \ \ \ \ \ \ a. Definition and Explanation\\ 
\ \ \ \ \ \ \ \ b. Classification\\ 
\ \ \ \ \ \ \ \ \ \ \ \ \ \ \ \ i. Binary Classification\\ 
\ \ \ \ \ \ \ \ \ \ \ \ \ \ \ \ ii. Multiclass Classification\\ 
\ \ \ \ \ \ \ \ c. Regression\\ 
\ \ \ \ \ \ \ \ \ \ \ \ \ \ \ \ i. Linear Regression\\ 
\ \ \ \ \ \ \ \ \ \ \ \ \ \ \ \ ii. Polynomial Regression\\ 
3. Unsupervised Learning\\ 
\ \ \ \ \ \ \ \ a. Definition and Explanation\\ 
\ \ \ \ \ \ \ \ b. Clustering\\ 
\ \ \ \ \ \ \ \ \ \ \ \ \ \ \ \ i. K-Means Clustering\\ 
\ \ \ \ \ \ \ \ \ \ \ \ \ \ \ \ ii. Hierarchical Clustering\\ 
\ \ \ \ \ \ \ \ c. Dimensionality Reduction\\ 
\ \ \ \ \ \ \ \ \ \ \ \ \ \ \ \ i. Principal Component Analysis (PCA)\\ 
\ \ \ \ \ \ \ \ \ \ \ \ \ \ \ \ ii. t-Distributed Stochastic Neighbor Embedding (t-SNE)\\ 
4. Evaluation and Validation\\ 
\ \ \ \ \ \ \ \ a. Training, Testing, and Validation Data\\ 
\ \ \ \ \ \ \ \ b. Accuracy, Precision, Recall, and F1-Score\\ 
\ \ \ \ \ \ \ \ c. Cross-Validation\\ 
5. Model Selection and Regularization\\ 
\ \ \ \ \ \ \ \ a. Bias-Variance Tradeoff\\ 
\ \ \ \ \ \ \ \ b. Overfitting and Underfitting\\ 
\ \ \ \ \ \ \ \ c. Regularization Techniques\\ 
% ...
6. Introduction to Neural Networks\\ 
...
        } \\
        \bottomrule
    \end{tabular}}
    
    \label{tab:case_course_plan}
\end{table*}


\section{Conclusion and Future Work}
This work is a pioneering exploration of an LLM-powered intelligent tutoring system, with an emphasis on the possibility of employing LLMs to complete complex and dynamic long-term interactions.
The proposed system, \modelname, can satisfactorily complete the core functions of an intelligent tutoring system. As ablation study shows, the three-process system design provides unique benefit in ensuring the stability and consistency of the system behavior, meanwhile maintaining flexibility and adaptiveness with the designed mechanism. 
Although our evaluation reveals the advantage of memory mechanism and process design in long-term interaction, we acknowledge that comprehensively evaluating an intelligent tutoring system is far more tricky~\cite{Graesser1995CollaborativeDP, cohen1982educational}. It is also important to design more standard metrics for interactive systems in the era of LLMs.
The system also faces concerns unique to LLMS, such as the validity of generated education content and the potential bias from training data~\cite{feng-etal-2023-pretraining, nozza-etal-2022-pipelines}, which might be mitigated by domain-specific fine-tuning and retrieval-augmented fact-checking.
Despite that, this work proposes a meaningful application of chaining LLMs in the educational process, which might inspire future efforts in employing LLMs to build more interactive and reflective systems.

\section*{Acknowledgement}
This work is supported by National Natural Science Foundation of China (Grant No. 62276154 \& No. 62236004), Research Center for Computer Network (Shenzhen) Ministry of Education, the Natural Science Foundation of Guangdong Province (Grant No. 2023A1515012914), Basic Research Fund of Shenzhen City (Grant No. JCYJ20210324120012033 and GJHZ202402183000101), the Major Key Project of PCL for Experiments and Applications (PCL2021A06), and the National Key R\&D Program of China (No.2022ZD0116312).

\bibliographystyle{ACM-Reference-Format}
\bibliography{custom}

\appendix

\end{document}